\documentclass[prb,12pt] {revtex4}
\usepackage[english]{babel}
\usepackage[intlimits]{amsmath}
\usepackage[dvips]{graphicx}
\usepackage{epsfig}

\begin{document}

\begin{titlepage}

\title{Translocation of  Polymers through Nanopores at Weak External Field, Direct
Approach with Monte Carlo Simulations}
\author{Stanislav Kotsev}
\email{kotsev@mpip-mainz.mpg.de}
\affiliation{Max-Planck-Institut f$\ddot{u}$r Polymerforschung, Ackermannweg 10, D-55128 Mainz, Germany}

\begin{abstract}
I did off-lattice $3D$ Monte Carlo simulations for polymer translocation 
through a narrow pore at low external field, trying to be as close 
to a direct approach as possible. The  process was found non 
equilibrium globally, but dynamics of the monomers close to the pore (the 
{\it fold}) was found close to quasi equilibrium. I observed a tension 
(or pressure) buildup  near the pore, and subdiffusion for the displacement 
of the middle monomer in the pore, as reported in the literature. I did 
not find the distribution of the reaction coordinate after some time of 
free diffusion predicted by the fractional diffusion equation.
\end{abstract}
\maketitle

\end{titlepage}

\section{Introduction}
Translocation of polymer molecules through nanopores is an
important process in biology.\cite{lodish_book} Examples are the
transport of DNA and RNA molecules
 across  pores of the nuclear membranes, injection of the virus
 genetic information, gene swapping etc.
Translocation of single-stranded DNA/RNA through a pore forced by
external voltage was demonstrated {\it in vitro} as a single
molecule signal by  Kasianowicz {\it et al.}
 \cite{kasianowicz96}. In such  experiment
voltage is applied across a membrane with a single pore in it. Due
to it negative charge, the DNA or RNA molecule is forced to drift
across (see the topical review by Meller \cite{meller_review}).
While inside the pore  the polymer blocks the ion current, thus
increasing the resistance, which effect can be detected with a single
event resolution. Such experiments showed  sensitivity to the
nucleotide composition, etc. \cite{meller_review}. It is now
believed that by using the blockage current only, no sequencing
can be achieved. Yet electrophoretic translocation through a hole
combined with another method is still one of the prime candidates
for fast DNA sequencing in the future (see the review by Zwolak \cite{zwolak}). 
Maybe most experiments use a pore made by self assembly of
 $\alpha$-hemolysin in  lipid bilayer. Such pores are so narrow 
(~1.4 nm diameter in the most narrow place), that only a single 
stranded DNA or RNA can pass. Typically the RNA/DNA is of the 
order of hundreds base units, the applied voltage is few hundreds  
mV (or less), and the blockage times are hundreds of microseconds. 
More recently experiments with silicone membranes have been 
reported. \cite{dekker_review} A big development in silicone 
nanopore production has been made in the last years. A pore of 1 nm 
diameter is reported, which similarly to the $\alpha$-hemolysin 
can translocate only single stranded DNA \cite{heng}. Obviously 
wider pores can be created and used, but this is outside the 
emphasis of this paper.

The translocation of a polymer through a narrow pore
driven by electric field is  also extensively simulated. Some authors
use MD simulations with simplified (e. g. cylinder like) pore
\cite{lansac, matysiak06, tian, randel, rabin05, fyta}, some consider
more detailed pore structure ($\alpha$-hemolysin)
\cite{muthukumar06, kong02, wells}. Another approach is dynamic Monte
Carlo simulations (only with simplified cylinder/square like pore)
\cite{chern01, loebl, milchev04, panja07, panja08, dubbeldamP,
dubbeldamE, wolterink}. The last approach allows the case without
external field to be simulated \cite{wolterink, panja07, panja08,
dubbeldamP}. No friction forces can be considered in a natural way
in Monte Carlo simulations, hence the hydrodynamic interactions
are neglected. Using some tricks that shorten the computer time,
MD simulations on the low external field problem have been recently
reported \cite{guillouzic, Wei}

The low external field behavior observed in simulations invoked  some
controversy. Considering equilibrium entropic barrier one gets for
the translocation time, reflecting boundary conditions at the
beginning  $\tau=C N^2$, where $N$ is the number of bases of the
polymer (or other parameter proportional to the length of the
chain). For ideal chain $C=\frac{\pi^2}{16 \tilde D}$ where
$\tilde D$ is $1D$ diffusion constant. In the above mentioned
result no friction is considered \cite{chuang01,
muthukumar99, sung96, slonkina03}. How noticed \cite{chuang01},  
the non forced translocation time has a simple lower limit - 
the time the polymer diffuses its radius of gyration (or  
translocation with no barrier). That gives  
$\tau \geq \frac{N^{2\nu}}{D}$, where $D$ is the diffusion 
constant of the whole molecule. In Rouse model $D \sim
\frac {1}{N}$ giving $\tau \geq  N^{2\nu+1}$. $\nu$ is the Flory
exponent, which in $3D$ is $\nu=0.5$ for ideal chain, and
$\nu=0.588$ for self avoiding random walk, corresponding to polymer
in a good solvent. The self avoiding case contradicts $\tau \sim N^2$.

Also, for long enough chains, the polymer shouldn't have time to
equilibrate at each step of it translocation. Equilibration time goes
as $n^{1+2\nu}$ with the number of monomers outside on one of the
sides of the membrane $n$. For ideal chain this gives $N^2$,
making equilibrium at each step  possible if not probable, and $N^{2.18}$
in the self avoiding case, resulting in equilibration slower than
translocation.

Extrapolating the result of $2D$ Monte Carlo simulations, a $\tau
\sim N^{1+2\nu}$ big $N$ behavior, good solvent was proposed 
\cite{chuang01,luo} (reflecting beginning, self avoiding chain, 
no external field). Later this result was supported in $3D$ by 
MD \cite{guillouzic, luo07, Wei}.  Based on $3D$ 
lattice \cite{panja08, wolterink, panja07}, or off lattice 
\cite{dubbeldamP} Monte Carlo much higher scaling exponents 
have been proposed for the above system, reaching up to \cite{panja07} 
$\tau \sim N^{2+\nu}=N^{2.588}$. Diffusion 
time scaling longer than squared length means subdiffusive behavior.

I performed off-lattice Monte Carlo simulations equivalent to
direct trial by transparent assumptions.  The whole dynamics of the 
process was obtained this way as transition rates, and then $\tau$ 
was calculated. I obtained for the above described system a smaller scaling 
exponent limited only by $\tau\geq N^{2\nu+1}$ in Rouse model. 

\section{The model}
I regard the mean translocation (or first passage) time $\tau$ of
polymer through a hole in a flat membrane. I start with the first
monomer passed, it cannot exit back (reflecting boundary
conditions), and finish when one monomer is left not translocated.
The membrane is thin only $\sim 1$ monomer is in the pore, the
others are outside. The hole is cylindrical and narrow, so the
chain cannot enter folded, and in the theoretical fit, the monomer
inside is assumed without degrees of freedom.

The polymer consists on $N$ beads. I use dynamic off-lattice
Monte Carlo model close to the bead-spring model \cite{milchev04,
gerroff, dubbeldamE, dubbeldamP}, which gives Rouse dynamics
\cite{gerroff}. Instead using springs, I just give some allowance
for the distance between the nearest neighboring monomers. For the
self avoidance I assume the bead is a hard sphere with radius
that prevents self crossing. I pick a monomer at random and
attempt to move it $\pm 0.5$ at each $x$, $y$, and $z$ direction
simultaneously with uniform distribution of the probability. The
move is rejected if it places the beads to close
 to (far from) one another or to the membrane, or accepted
otherwise. This is equivalent to $0$ or $\infty$ potential, or
there are  no attractive or repulsive interactions. $N$ such
attempts are the unit of time (one MC step).
 No external field is applied - all the motion is completely
stochastic by nature.

I regard self avoiding and phantom chains. For the self avoiding case
the distance between two nearest neighbors is between $rmin=0.5$,
 and $rmax=0.7$. With respect to not nearest neighbors, and the
membrane the bead is a hard sphere with radius  $0.3$, which makes
the self crossing of the chain practically impossible. The
membrane is $0.11$ in thickness, so no additional care is needed
for the polymer not to drift with no bead in the pore. A thicker
membrane slows down the translocation. The pore is cylindrical
with radius $0.4$ -just a little bit bigger than the bead.

For the phantom polymers only nearest neighbors distance is
controlled, self crossing is allowed. The distance between two
nearest neighbors is again between $rmin=0.5$, and $rmax=0.7$. The
membrane is $0.71$ in thickness for the same reasons as above. The
pore is cylindrical with radius $0.1$.

I checked the two models by using  a polymer of $n$ monomers with
the first one  fixed on the membrane surface. $\vec{r}$ connects
the first and the last bead. First I compared the mean of the component of 
$\vec{r}$ in direction perpendicular to the membrane. Similarly to
the bulk case this should give scaling as $n^{\nu}$. I also
checked the equilibration time of such chain defined as the
correlation coefficient (\ref{correleator}) between $\vec{r}$ and
it initial position as a function of the time. As mentioned one
expects scaling as $n^{1+2\nu}$. I did this for chains of $n=294$
and $n=6$ monomers for each model. For the phantom case I got
$\nu\sim 0.5$ within our accuracy from the both tests, as
expected. For the self avoiding case I got $\nu\sim0.6$ from the
distance test, and $n^{2.3}$ as an approximation of $n^{1+2\nu}$
from the correlation test.  Thus I overestimated a little bit the
theoretical expectation of $\nu=0.588$. Note that for similar
polymer lengths similar scaling is reported for dynamical exponent
\cite{gerroff} as a reasonably close to the theoretical one (which
should be observed when $n \to \infty$). The obtained exponents
proof that the polymer is effectively in a good solvent.

The number of translocated monomers is the reaction coordinate
$k$. I actually simulate the effective rates for the polymer to
jump $step$ monomer forwards or backwards. Then one can calculate
the translocation time  $\tau$ from (\ref{Pury}) and (\ref{prob}).
I use $step=6$ monomers (unless otherwise specified). This is
much (about a order of magnitude for the regarded systems) faster than to
wait for a direct escape starting from the beginning. I start
from $k=\tilde{k}$ and simulate till the chain reaches both  $k=\tilde{n} +
step$ and $k=\tilde{k} - step$ at least once. Than I start from
$k= \tilde{n} + step$. The program keeps the configuration of the polymer
when first reached translocation to $k=\tilde{k} + step$ and
use it as the initial state when starting from the new $k=
\tilde{k} + step$ state. I use (\ref{ansatz}) to calculate the
effective rates $u_m^{effective}$ and $w_m^{effective}$.

I did 500 or more trials  for every simulated point shown,
unless otherwise mentioned.

I define $k=\tilde{k} \pm step$ or $\tilde{k}$ when the
corresponding monomer is at the middle along it translocation in
the pore. In the off-lattice MC monomers jump randomly, which
introduce some error in defining the position. Reducing the
relative error, a bigger $step$ makes the result more accurate,
but demands longer computer time.

Please note, that in this work I regard the mean translocation
time. This is the case for most theoretical works and simulations. 
In experimental works the most probable passage time is
measured. For theoretical works dealing with the shape of the
translocation time distribution (not zero external field) see
\cite{metzler03, lubensky99, flomenbom03}.

\section{Theoretical approach}
I use the formula for the mean time $\tau$ to first reach $b$ or
$e$. Let's  $b$ is before $e$, and  we start from $s$ between $b$
and $e$. For discrete $1D$ walk \cite{pury03,vanKampen}:
\begin{eqnarray}\label{Pury}
\tau (b-1,s,e+1)&=&
\pi (b-1,s,e+1) \left( \sum_{k=b}^{e} \frac{1}{u_k} +  \sum_{k=b}^{e-1} \frac{1}{u_k} \sum_{i=k+1}^{e} \prod_{j=k+1}^{i} \frac{w_j}{u_j}         \right) \nonumber \\
&-& \left( \sum_{k=b}^{s-1} \frac{1}{u_k} +  \sum_{k=b}^{s-2}
\frac{1}{u_k} \sum_{i=k+1}^{s-1} \prod_{j=k+1}^{i} \frac{w_j}{u_j}
\right)
\end{eqnarray}
where
\begin{equation}\label{prob}
\pi (b-1,s,e+1)=\frac{1+\sum_{k=b}^{s-1} \prod_{j=b}^{k}
\frac{w_j}{u_j}}{1+\sum_{k=b}^{e} \prod_{j=b}^{k} \frac{w_j}{u_j}}
\end{equation}

is the probability starting at $s$ to reach  $e+1$ before ever
reaching $b-1$. Above $u_k$ is the rate per unit time to jump
forwards from $k$ to $k+1$. $w_k$ is the same  but backwards from
$k$ to $k-1$.

We also need the ansatz (see Fig. 1)
\begin{eqnarray}\label{ansatz}
&& u_m^{effective}=\frac{\pi (b_m,s_m,e_m)}{\tau (b_m,s_m,e_m)} \nonumber \\
&& w_m^{effective}=\frac{1-\pi (b_m,s_m,e_m)}{\tau (b_m,s_m,e_m)} \nonumber \\
&& e_m=s_{m+1};  \quad b_m=s_{m-1}
\end{eqnarray}
the last is consistent with the formula for mean first passage time,
or if one uses $u_m^{effective}$, and $w_m^{effective}$ instead of
$u_k$, and $w_k$ in Eqs.(\ref{Pury}), one gets the same time. Note
that the set of $u_m^{effective}$, and $w_m^{effective}$ should be
less in number than the one of $u_k$, and $w_k$  in order
(\ref{ansatz}) to be needed.

To estimate the rates theoretically I use:
\begin{eqnarray}\label{coeff}
u_k=\frac{\tilde D}{\Delta^2} \exp \left( \frac{F(k)-F(k+1)}{2 k_B T} \right) = \frac{\tilde D}{\Delta^2} \sqrt{\frac{Z_{k+1}}{Z_k}}\nonumber \\
w_k=\frac{\tilde D}{\Delta^2} \exp \left( \frac{F(k)-F(k-1)}{2 k_B T} \right) =\frac{\tilde D}{\Delta^2} \sqrt{\frac{Z_{k-1}}{Z_k}}
\end{eqnarray}
where $\tilde D$ is the $1D$ diffusion constant if the distance 
traveled for a jump between 2 neighboring states is $\Delta$. 
$F(k)$ is the free energy at state $k$.  The factor of $2$ in 
denominator insures the detailed balance between the neighboring states.

One may assume a smooth continuous along $k$ potential and put
additional states between the original ones, shortening $\Delta$. 
One sees from (\ref{ansatz}) that $\tilde D $ should be kept
constant to reproduce $u_m^{effective}$ and $w_m^{effective}$ 
between the original states the same as the original rates if 
$F$ is constant, and  $\Delta \to 0$. This way one obtains from 
(\ref{coeff}) and (\ref{Pury}) the continuous  limit for the 
case we start from state $0$, with reflecting boundary conditions 
there ($w_0=0$):

\begin{equation}\label{continuous}
\tau(0,0,N)_{w_0=0}=
\int_{0}^{N} dk \exp \left( \frac{F(k)}{k_B T} \right)
\int_{0}^{k} di \frac{1}{\tilde D(i )}\exp \left(- \frac{F(i)}{k_B
T} \right)
\end{equation}

For polymer of $n$ segments attached at one of its end to a flat surface
 the partition function is \cite{eisenriegler}:
\begin{equation}\label{attached_chain}
Z_n \sim n^{\gamma \prime -1}
\end{equation}
Hence the entropic potential of our system has the form
\begin{equation}\label{potential}
\frac{F(k)}{k_B T}=(1-\gamma \prime) \ln(k) + (1-\gamma \prime)
\ln(N-M-k) + const
\end{equation}
In the last equation $M$ is the number of the monomers inside the
pore, which is assumed without degrees of freedom. $k$ is the
number of translocated monomers or our reaction coordinate, and:
\begin{equation}\label{gamma_prime}
\gamma \prime= \left\{\begin{array}{l}{\displaystyle
{1 \over 2};\quad} \\
\\{\displaystyle 0.69;\quad}
 \\
\\{\displaystyle 1;\quad}
\end{array}\right.   \begin{array}{l} {\rm phantom ~chain}\\
\\ {\rm self~ avoiding ~chain}\\
\\ {\rm rigid ~rod}
\end{array}
\end{equation}

According to (\ref{coeff}) and (\ref{potential}) our system have
the symmetry:
\begin{eqnarray}\label{symmetry}
u_k= w_{N-M-k}\nonumber \\
w_k= u_{N-M-k}
\end{eqnarray}

How we know\cite{chuang01,muthukumar99, sung96, slonkina03}
(\ref{potential}) with (\ref{continuous}) gives $\tau
(0,0,N-M)_{w_0=0}=C (N-M)^2$ with  $C=\frac{\pi^2}{16 \tilde D}$
for ideal chain. 

\section{How equilibrium is the translocation process?}
I first use the data for phantom chain. How noted, the translocation
process for them is expected to be closer to quasi equilibrium one. 
All the simulated data shown are fitted to the theory by one constant.
I calculate and simulate the first exit time for the polymer
starting with one base passed, which cannot reenter back - $w_1=0$
(reflecting boundary conditions), and finish at one monomer left -
$\tau (0,1,N-1-M)_{w_1=0}$. In all the data $M=1$.
Note that in this case the forward reaching probability $\pi =1$ 
in (\ref{Pury}).

In  Fig. 2  is shown a fit of the effective forward rate
$u_m^{effective}$ for ideal polymer to jump $step=6$  monomers
forwards, $N=303$. The theoretical values are calculated using
(\ref{coeff}), (\ref{potential}) and then (\ref{ansatz}) to get
the effective value for $step=6$ monomers. One sees almost perfect
agreement with quasi equilibrium theory. The simulated data for
$\tau (0,1,N-1-M)_{w_1=0}$  as function of the number of the
monomers in the chain - $N$ are shown in the Fig. 3 for the case
of ideal chain. Since the rates agree very well with theory,  I
simulate only half of them, than use their symmetry
(\ref{symmetry}) to find the rest (and $\tau (0,1,N-1-M)_{w_1=0}$
). The slope goes down with $N$, reaching  $N^{2.2}$.
In this graph one sees again a very good agreement with the
quasi equilibrium discrete process theory, given by (\ref{coeff}),
(\ref{potential}), and (\ref{Pury}). I believe for bigger $N$ the
exponent will drop to $2$.

The next question is if these data correspond to
quasi equilibrium process. To estimate a lower limit of the
equilibration time of a chain of $n$ monomers anchored to the
pore at one end, I use the correlation coefficient $c(n,t)$:
\begin{equation}\label{correleator}
c (n,t)=\Big\langle\frac{(\vec{r}(t)-\langle \vec{r}\rangle).(\vec{r}(0)-\langle \vec{r}\rangle)}{\sqrt{(\vec{r}(t)-\langle \vec{r}\rangle)^2(\vec{r}(0)-\langle \vec{r}\rangle)^2}}\Big\rangle
\end{equation}

In the above definition $\vec{r}$ is the vector from the
end of the polymer fixed in the membrane (the pore) to it free end, and
$\langle \quad \rangle$ means averaging over ensemble or over time
which I assume equivalent. $c(n,t)$ is correlation coefficient of
the direction of $\vec{r}$ , hence it should vanish the same as or 
faster than the real memory of the whole chain. I did simulations for
anchored chains of $n=294$ and $n=6$ monomers for both self
avoiding and phantom models. The observed behavior is
$c(n,t)=\exp(-u\frac{t}{n^\omega})$ where for $n \to \infty$ one
should have $\omega=1+2\nu$, and $u$ is specific for each model.
As noted for the phantom case I really have the big $n$ value
$\omega=2$, while for the self avoiding model $\omega=2.3$.
Similar approach and results with slightly different correlation
coefficient, and using lattice dynamics are reported
\cite{panja08} for the same system. For the phantom case I show 
the times $t_e(n)$ and $t_{10}(n)$, (where $c(n,t_e)=\frac{1}{e}$, and
$c(n,t_{10})=\frac{1}{10}$) in Fig. 4. Together are the mean time
for the polymer to move $step=6$, monomers forwards or backwards
which are directly simulated, as well the mean lifetime the polymer
to move one monomer left or right (which come as $\frac{1}{w+u}$
from the theoretical values   from (\ref{coeff}), and
(\ref{potential}), also used to get the theoretical fit in Fig.
2). One sees that (with probable exception of near the center),
there was not been a real equilibrium.

To see why in such case the data look like equilibrium again the
phantom chain is used. In Fig.5  effective forward rate 
$u_m^{effective}$ is given for the polymer to jump $step=6$  monomers 
forwards, $N=303$, and $N=807$. This is the very beginning of the 
process ($k$ close to $0$). I also give the theoretical prediction 
for the rate with the same translocated number of polymers, but 
replacing the rest (non translocated end) of the chain with much 
shorter or longer one (total number becomes $N=40$, or $N=4000$). 
One sees, that this doesn't change much the translocation rate. 
Or, if the monomers on one of the sides of the membrane have a 
counter part on the other of few tens of monomers or of few 
hundreds, this doesn't change much the dynamics around the pore. 
It worths to mention, that (\ref{coeff}) with (\ref{potential}) 
saturates to limiting values for the beginning  of the process 
(or by  (\ref{symmetry}) the end of the process) as $N$ grows:
\begin{eqnarray}\label{coeff_limit}
u_k&=&\frac{\tilde D}{\Delta^2} \left(\frac{k}{k+1} \right)^{\frac{1-\gamma \prime}{2}} \nonumber \\
w_k&=&\frac{\tilde D}{\Delta^2} \left(\frac{k}{k-1} \right)^{\frac{1-\gamma \prime}{2}} \nonumber \\
k &-&{\rm small}, \quad N \to \infty
\end{eqnarray}
So, I believe that the process is equilibrium for chains of few tens 
of monomers. In case of bigger chains, the ones close to the pore do 
not feel the rest of the chain more than few tens of monomers apart. 
In other words the rates are close to ones if everything greater 
than few tens monomers each side of the pore is cut. How suggested 
\cite{dubbeldamP, dubbeldamE} there is a partial equilibrium for 
the few tens of monomers close to the pore.

In Fig. 5 one also sees that  $u_1^{effective}$ is consistently lower than
the theoretical prediction. I interpret this fact as that there was a
local equilibrium, and the rest of the monomers (not being equilibrated),
just pull slightly backwards.

\section{Does the self avoidance keeps the theory?}
Next I present data for self avoiding chain. Again they all are
fitted with one constant. Due to the different lengths of the pores
(thicknesses of the membranes), this constant is incompatible to
the one for phantom chain.

In Fig. 6 are given the the effective forward rate
$u_m^{effective}$ for the polymer to jump $step=6$  monomers
forwards for $N=75$, and the same for $step=30$  monomers forwards
for $N=123$. One may see that for the larger chain, there is some
deviation from the theory. That is why I do not use
(\ref{symmetry}), but rather simulate all the rates needed to
calculate the translocation time. For the same reason, I also
give in Fig. 7 times obtained by direct trials. One may observe
the nearly perfect match between the theory, and the two ways of
simulation. The translocation time $\tau (0,1,N-1-M)_{w_1=0}$, $M=1$ 
ends up scaling as $N^2$. I conclude, that the hypothesis for 
local equilibrium next to the pore is still valid.

What violates slightly the theory is that the number of
translocated monomers is not a perfect reaction coordinate. It
doesn't give all the information for the state of system, namely 
that the polymer is stretched when it just advances (see the next 
section).

\section{Short time dynamics of self avoiding polymer}

The procedure described above (but with $step=2$ which 
shortens the computation time) is used to thread a self avoiding polymer 
exactly to the position where the middle monomer is in the pore. 
Then I set the time to zero, and observe the further dynamics. 
The entropic potential is then zero, and doesn't affect the process. 

It is believed that  for small enough times  the dynamics is 
subdiffusive. The data for  $SD^2$ of the displacement of 
the middle monomer along the pore really scale as $t^{0.7}$ 
for the range shown in Fig. 8, $N=127$. Interestingly the 
$SD^2$ for the reaction coordinate $k$ scales linearly for 
this range. The displacement of the middle  monomer along 
the pore also scales differently than the change of $k$.  
The author explains this with the stretching of the chain. 
Unfortunately very short time intervals are hardly accessible.
In the off lattice Monte Carlo monomers jump to random 
positions, and for small times the error in defining $k$ will
become compatible to it value.

Within the accuracy, there is no deviation from Gaussian 
distribution of the reaction coordinate $k$ at some fixed time 
after starting  from the middle (for the range shown in 
Fig. 8). The last is in agreement with the lattice simulations 
described \cite{panja08} where they have observed minute 
deviation far from the mean value for small times. This 
is also in agreement with the simulations  \cite{kantor}, 
for phantom polymer in $1D$. Deviation from Gaussian 
distribution in the middle - a non smooth cusp is predicted 
by the fractional diffusion equation, describing subdiffusion 
with diverging mean waiting times between the consecutive 
moves. So, my results are in parallel with the statement 
 \cite{kantor} that for the polymer translocation we 
may have a different type of subdiffusion.

The average shift of the reaction coordinate $k$ is shown 
in Fig. 8. One sees that the average position consistently 
goes back. That is why we didn't get perfect agreement with the 
theory for $N=123$. The reason is the accumulated tension in the 
chain. My results for the mean displacement of $k$ due to the 
relaxation of the over stretching confirm the tension 
relaxation exponent $\alpha=\frac{1+\nu}{1+2\nu}$  introduced 
in \cite{panja07}. The evolution of the $SD(k)$ is completely 
diffusive for the times shown in Fig. 8 within our 
accuracy.  In parallel with what I said the dynamics 
is again local, $SD$ of  $k$ is practically independent on $N$ -
see Table 1. The data for the average shift $k$ shown 
in Fig. 8 are less independent on $N$. 
	
The author speculates that such dependence of the dynamics from
the history is the explanation of the subdiffusive behavior 
expected for longer chains.

\section{What is the long chain translocation time scaling? }

The last point is to check why the exponent is lower than the
theoretical limit   $\tau \geq  N^{2\nu+1}$ for the Rouse dynamics.
I performed a direct simulations (200 trials) for the self avoiding 
polymer using pore much bigger than the polymer length (or
practically absent). I did trials for $N=125$ and $N=175$. In 
agreement with our equilibration times (using  the correlation 
coefficient (\ref{correleator})) the time scales as $N^{2.3}$.
Comparing them with the self avoiding data for normal pore shown
on Fig. 7 they are $\sim 50$ times smaller. I estimated, that
following such behavior this time will surpass the one for the 
regular data for $N\sim 10^{12}$. In other words the data are too 
far from the point where the the scaling $N^{1+2\nu}$ will become important.
Let's note that it is hard to extrapolate the same physics for
 such big $N$. On the other hand, all the simulation data 
(including the ones presented here ) start with larger exponent, 
which goes down with increasing the $N$. Here an exponent 
lower than the limit for the Rouse model is reached, which suggests that
 $\tau \sim \frac{N^{2\nu}}{D}$ is the real exponent. Also, Rouse
model (and hence MC) overestimates the hydraulic friction in $D$.
 For the real systems, $D \sim \frac {1}{N^{\nu}}$ (Zimm exponent) 
which doesn't contradict $\tau \sim N^2$. For arguments that the Zimm exponent
may not be valid for our problem, see \cite{guillouzic, chuang01}.

\section{Acknowledgments}
The author is grateful  to Anatoly B. Kolomeisky for the 
support and for the useful discussions.

\newpage

\begin{table}[ht]
\begin{center}
\begin{ruledtabular}
\begin{tabular}{c c c c}
 time  & $N=127$ &$N=179$ & $N=299$ \\ 
MC steps\\
\hline
$10^4$ & 0.80 & 0.79 & 0.76 \\
\hline
$10^5$ & 2.6 & 2.6 & 2.7  \\
\hline
$10^6$ & 8.3 & 8.2 & 7.8  \\

\end{tabular}
\end{ruledtabular}
\end{center}
\end{table}

\vskip 5in

\noindent Table 1.  \quad The $SD(k)$ of a self avoiding polymer with time, 
due only to diffusion. $SD(k)=0$ at $t=0$.

\newpage

\noindent {\bf Figure Captions:} \\

\noindent Fig 1.  \quad Using the effective rates $u^{effective}$ and
$w^{effective}$ turns the system into equivalent one, consisting of
smaller number coarse grained states.

\vspace{5mm}

\noindent Fig 2.  \quad  Simulated and theoretical rates
$u^{effective}$ , $step=6$  as function of the number of the monomers passed.
The theoretical fit is done using quasi equilibrium approach.

\vspace{5mm}

\noindent Fig 3.  \quad Translocation time $\tau
$ as a function of the length of the chain $N$.
One monomer stays in the pore, process starts with one base passed,
which cannot reenter back - $w_1=0$ (reflecting boundary
conditions), and finish at one monomer left. The theoretical fit
is done using quasi equilibrium theory for discrete states.

\vspace{5mm}

\noindent Fig 4.  \quad Equilibration time at one of the sides 
of the pore (thin lines) and the time to pass $step$ monomers 
during spontaneous translocation (thick
lines). Upper thin lines is the time the correlation coefficient 
(\ref{correleator}) to vanish down to $c(n,t)=\frac{1}{10}$,
the lower thin line - the same for  $c(n,t)=\frac{1}{e}$.
Upper thick line is the average time the polymer of $N=303$ needs
to move $step=\pm6$ monomers which is simulated to get
$u^{effective}$ and $w^{effective}$, lower thick line - the 
same for $step=1$. The diamonds are for $N=807$ and $ step=\pm6$.

\vspace{5mm}

\noindent Fig 5.  \quad First few rates $u^{effective}$ , $step=6$ 
for big phantom chains. The lines are calculated, as the original
 polymer is replaced with one having much shorter non translocated 
tail (total length $N=40$, upper curve), or much longer non 
translocated tail (total length $N=4000$, lower  curve). Quasi 
equilibrium discrete theory is used.

\vspace{5mm}

\noindent Fig 6.  \quad Simulated and theoretical rates
$u^{effective}$ as function of the number of the monomers passed.
The
theoretical fit is done using quasi equilibrium approach.
In the case of $N=75$, $step=6$ and the fit is good. In the case $N=123$
$step=30$, and we have performed only $200$ trials.  In the last  data one sees
a consistent deviation from the theory, due to the accumulation of
tension in the chain. That is the reason that $step=6$ cannot be used.

 \vspace{5mm}

\noindent Fig 7.  \quad Translocation time $\tau
$ as a function of the 
length of the chain $N$.
One monomer stays in the pore, process starts with one base passed,
which cannot reenter back - $w_1=0$ (reflecting boundary
conditions), and finish at one monomer is left. The theoretical fit
is done using quasi equilibrium theory for discrete states. 3 of
the times are simulated using average of $200$ direct trials.

\vspace{5mm}

\noindent Fig 8.  \quad The polymer was threaded in using the
procedure described in the text, $step=2$,
equivalent to spontaneous translocation.  When the middle is
reached, the further dynamics is observed, which is due
entirely to the accumulated tension. The line corresponds to
$t^{\alpha}$ with tension relaxation exponent
$\alpha=\frac{1+\nu}{1+2\nu}$ as predicted by \cite{panja07}.

\newpage

\begin{figure}[ht]
\begin{center}
\unitlength 1in
\begin{picture}(4.0,4.0)
  \resizebox{3.0in}{3.0in}{\includegraphics{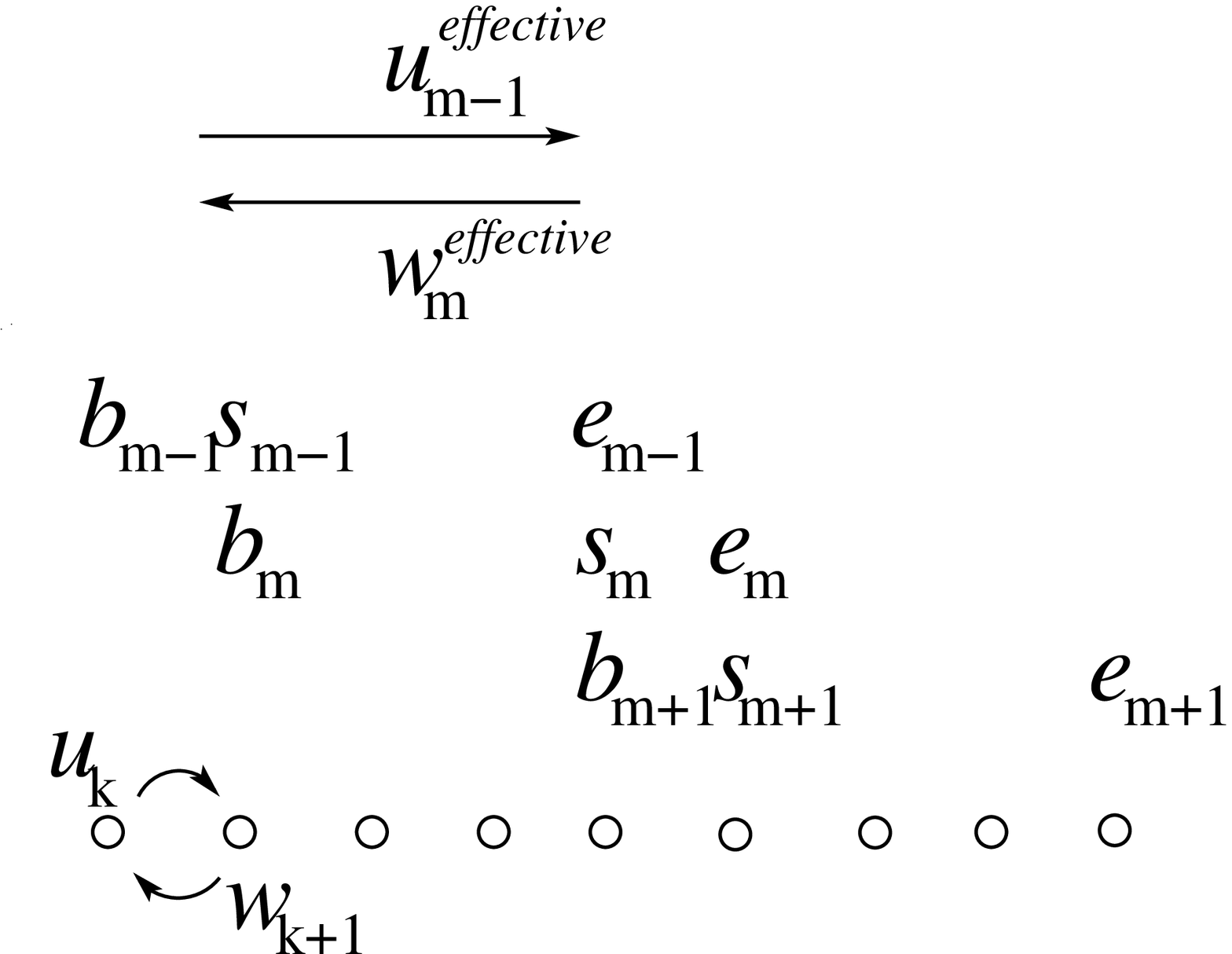}}
\end{picture}
\vskip 3in
 \begin{Large} Figure 1. Kotsev \end{Large}
\end{center}
\end{figure}

\newpage

\begin{figure}[ht]
\begin{center}
\unitlength 1in
\begin{picture}(4.0,4.0)
  \resizebox{3.0in}{3.0in}{\includegraphics{coefftheory303.eps}}
\end{picture}
\vskip 3in
 \begin{Large} Figure 2. Kotsev \end{Large}
\end{center}
\end{figure}

\newpage

\begin{figure}[ht]
\begin{center}
\unitlength 1in
\begin{picture}(4.0,4.0)
  \resizebox{3.0in}{3.0in}{\includegraphics{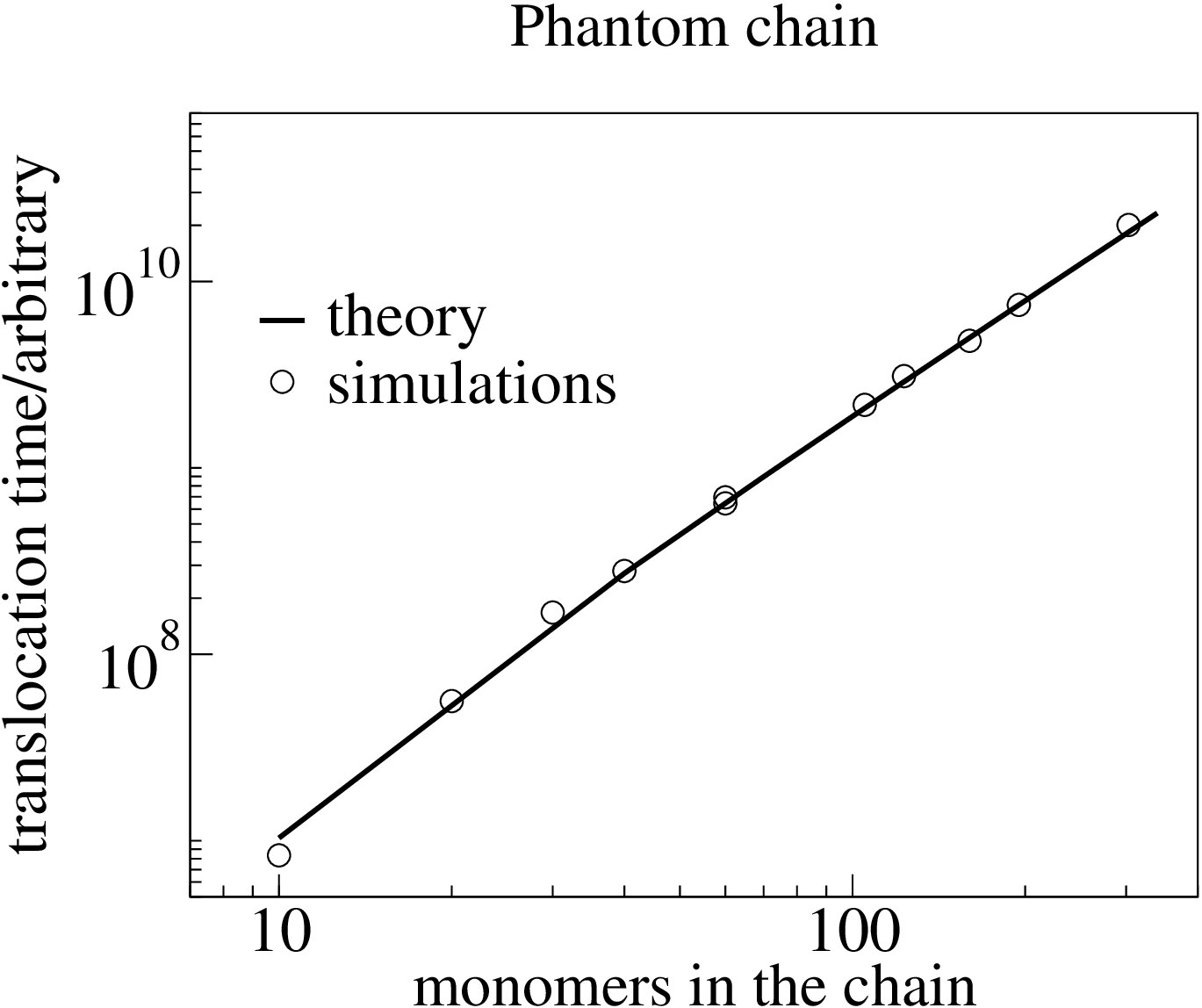}}
\end{picture}
\vskip 3in
 \begin{Large} Figure 3. Kotsev \end{Large}
\end{center}
\end{figure}
\newpage

\begin{figure}[ht]
\begin{center}
\unitlength 1in
\begin{picture}(4.0,4.0)
  \resizebox{3.0in}{3.0in}{\includegraphics{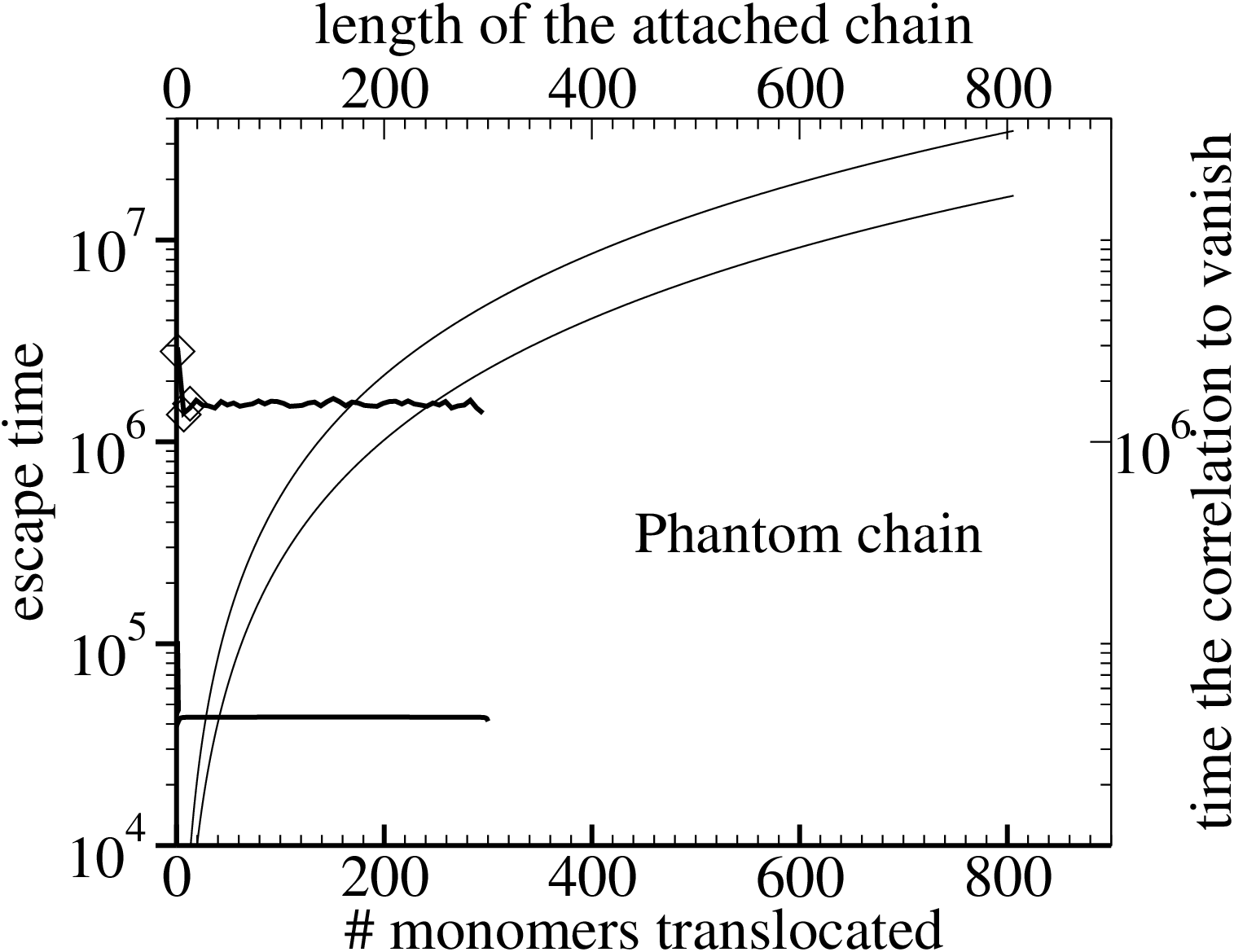}}
\end{picture}
\vskip 3in
 \begin{Large} Figure 4. Kotsev \end{Large}
\end{center}
\end{figure}
\newpage

\begin{figure}[ht]
\begin{center}
\unitlength 1in
\begin{picture}(4.0,4.0)
  \resizebox{3.0in}{3.0in}{\includegraphics{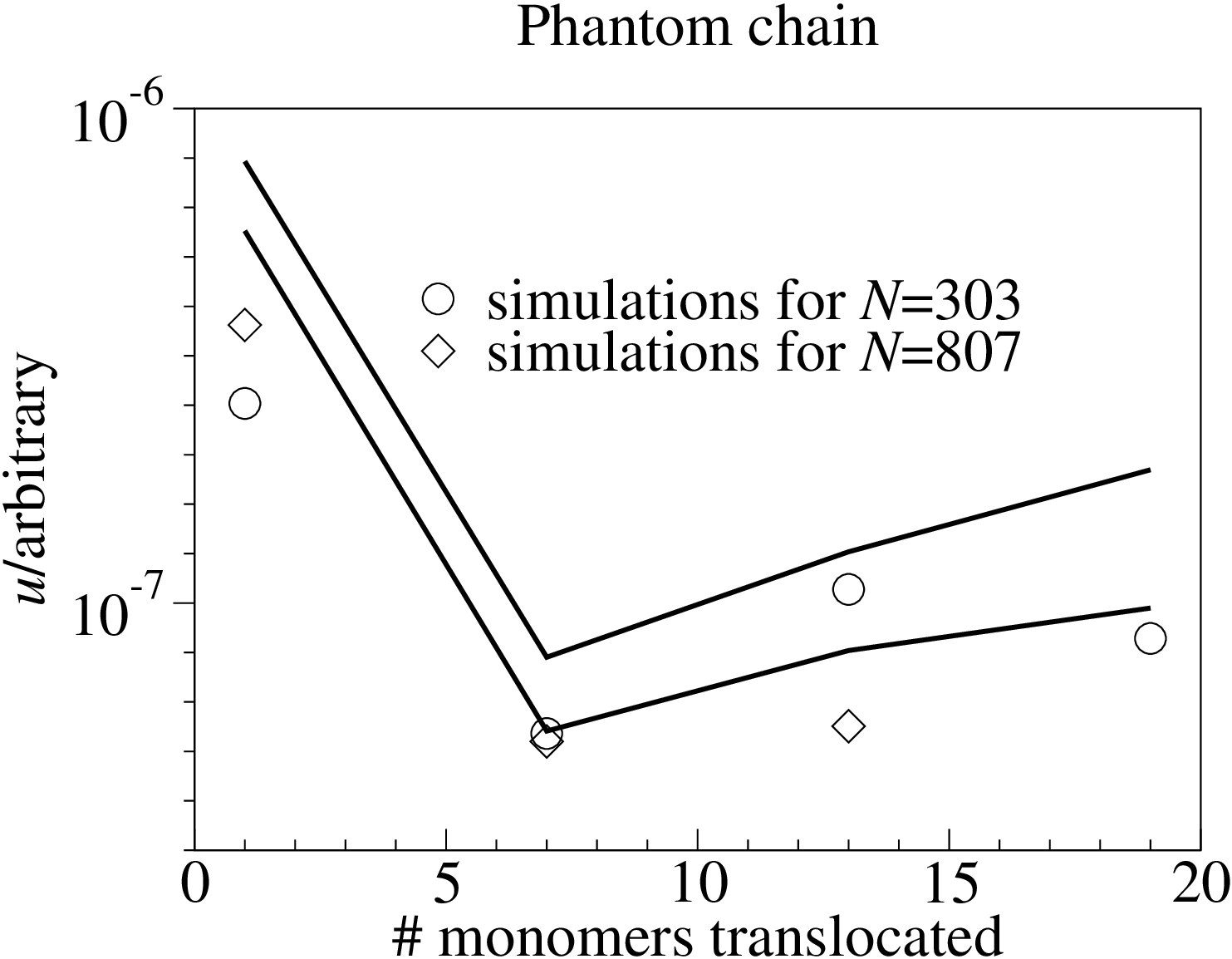}}
\end{picture}
\vskip 3in
 \begin{Large} Figure 5. Kotsev \end{Large}
\end{center}
\end{figure}
\newpage

\begin{figure}[ht]
\begin{center}
\unitlength 1in
\begin{picture}(4.0,4.0)
  \resizebox{3.0in}{3.0in}{\includegraphics{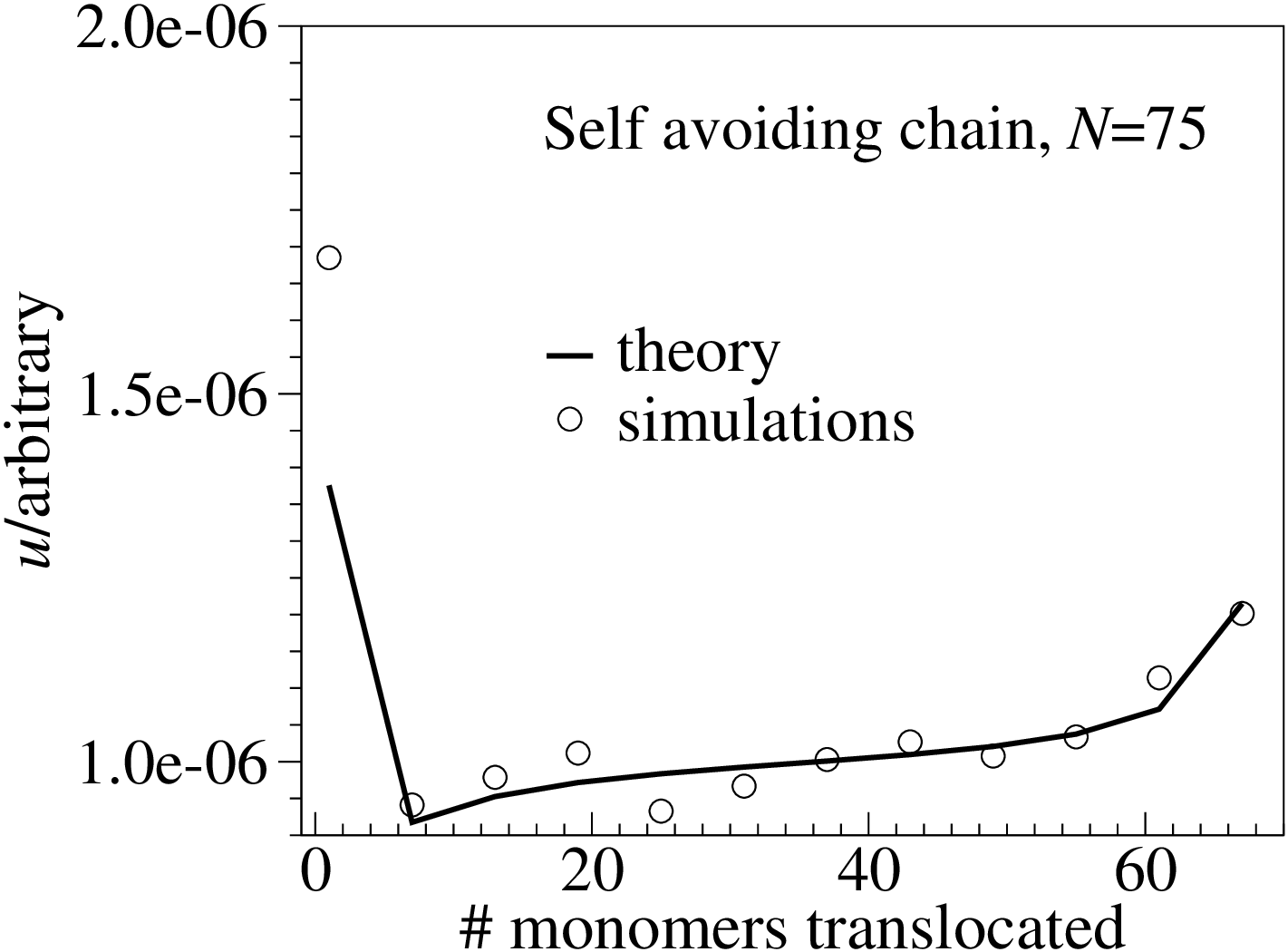}}
\end{picture}
\begin{picture}(4.0,4.0)
  \resizebox{3.0in}{3.0in}{\includegraphics{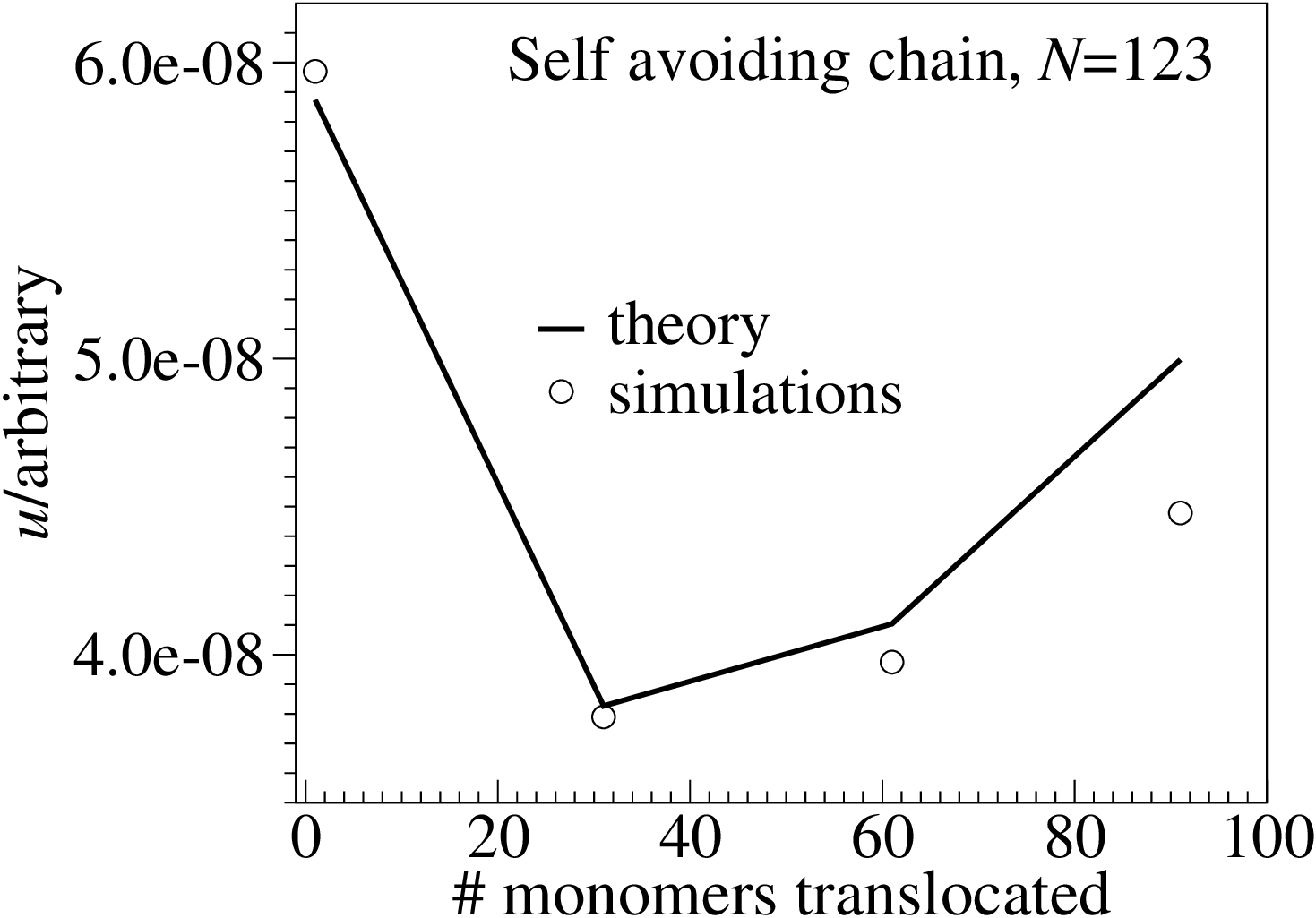}}
\end{picture}
\vskip 1.5in
 \begin{Large} Figure 6. Kotsev \end{Large}
\end{center}
\end{figure}

\begin{figure}[ht]
\begin{center}
\unitlength 1in
\begin{picture}(4.0,4.0)
  \resizebox{3.0in}{3.0in}{\includegraphics{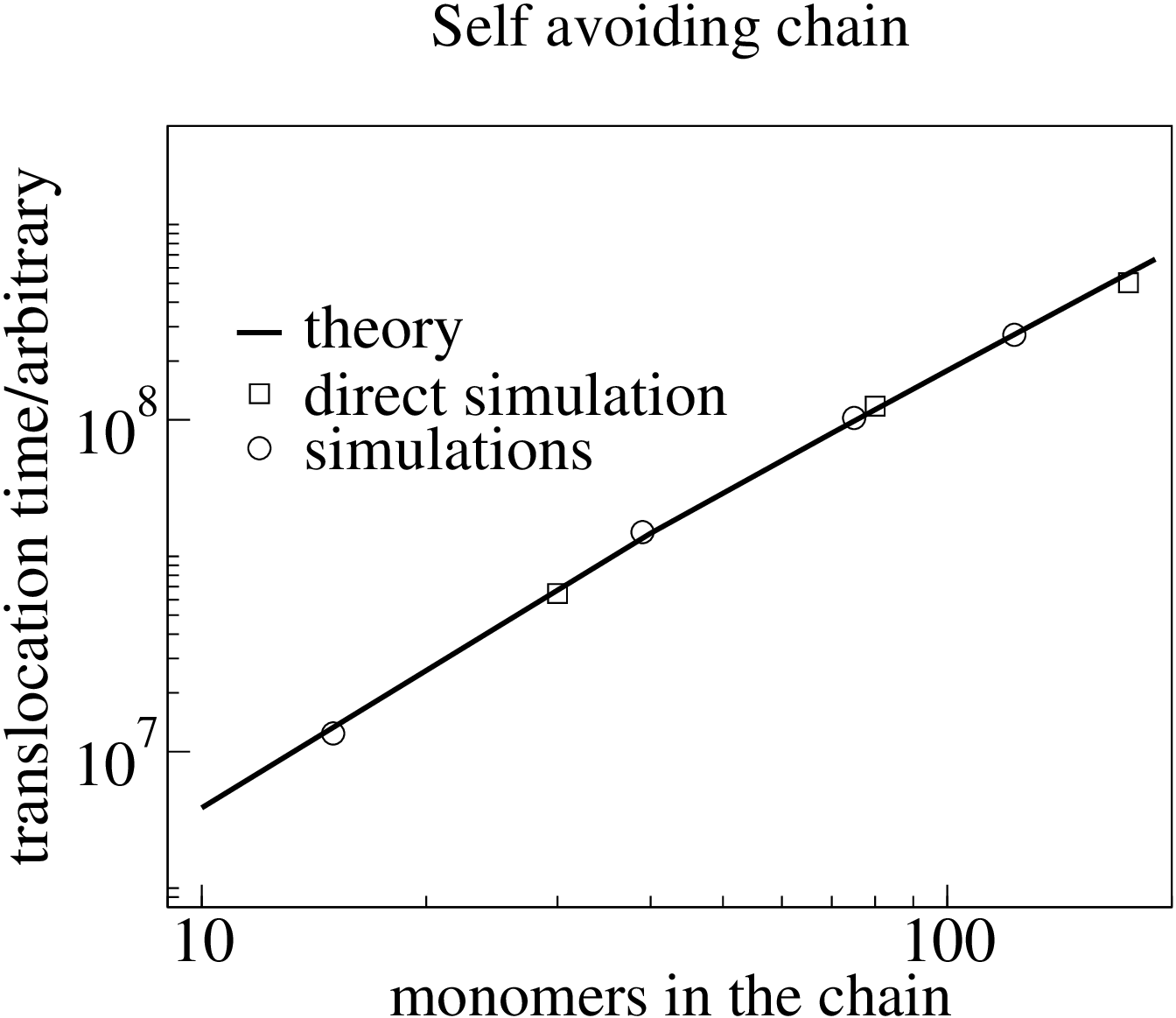}}
\end{picture}
\vskip 3in
 \begin{Large} Figure 7. Kotsev \end{Large}
\end{center}
\end{figure}
\newpage

\begin{figure}[ht]
\begin{center}
\unitlength 1in
\begin{picture}(4.0,4.0)
  \resizebox{3.0in}{3.0in}{\includegraphics{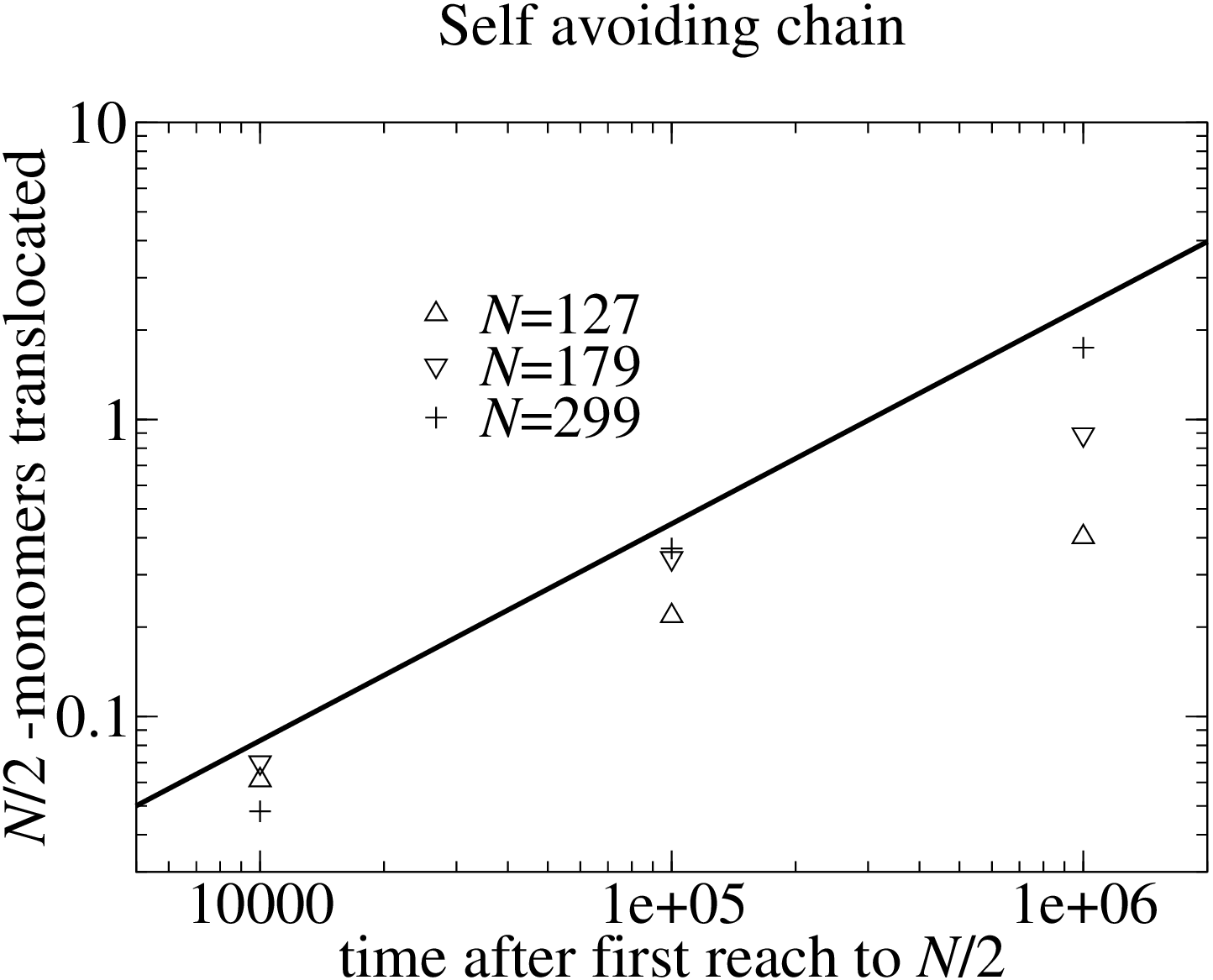}}
\end{picture}
\vskip 3in
 \begin{Large} Figure 8. Kotsev \end{Large}
\end{center}
\end{figure}
\newpage

\end{document}